\documentstyle[aps,twocolumn]{revtex}
\begin{document}
\draft
\title{Interference Phenomenon for Different Chiral Bosonization Schemes}
\author{Everton M. C. Abreu and Clovis Wotzasek}
\address{Instituto de F\'\i sica, UFRJ\\
	Rio de Janeiro, RJ, Brazil}
\date{\today}
\maketitle
\begin{abstract}
\noindent We study, in the framework put forward by Siegel\cite{WS}
and by Floreanini and Jackiw\cite{FJ} (FJ), the relationship between different chiral bosonization schemes (CBS). 
This is done in the context of the soldering formalism\cite{MS}, that considers
the phenomenon of interference in the quantum field theory\cite{ABW}.
We propose a field redefinition
that discloses the presence of a noton, a nonmover field, in Siegel's
formulation for chiral bosons.
The presence of a noton in the Siegel CBS is a new and surprising result,
that separates dynamics from symmetry.
While the FJ component describes the dynamics, it is the noton that carries
the symmetry contents, acquiring dynamics upon quantization
and is fully responsible for the Siegel anomaly.
The diagonal representation proposed here is used to study the effect
of quantum interference between gauged rightons and leftons. 
\end{abstract}


\noindent {\bf 1} Recently\cite{ABW} a new interpretation 
for the dynamical mass generation known as Schwinger mechanism,
has been proposed.  This study, made
in the context of the soldering formalism\cite{MS}, that considers the interference of
right and left gauged FJ chiral bosons\cite{FJ}.
The result of the chiral interference shows the presence of a massive
vectorial mode, for the special case
where the Jackiw-Rajaraman regularization parameter is $a=1$\cite{JR}, which is the value where the chiral theories have only one massless excitation in the spectrum.
This clearly shows that the massive vector mode results from
the interference between two massless modes.

Incidentally, the FJ model is the 
gauge fixed version of the more general CBS
proposed by Siegel\cite{WS}.  
The Siegel modes (rightons and leftons)
carry not only
chiral dynamics but also symmetry information. 
The symmetry content of the theory is well
described by the Siegel algebra, a truncate
diffeomorphism, that disappears at the quantum level.
The chiral interference between gauged rightons and leftons should lead to
a massive vector mode, plus the symmetry of
the combined right-left Siegel algebras.
As shown in the section II, the result of the chiral interference between gauged
Siegel modes is a Hull noton
\cite{CH}. This represents only the symmetry part of the
expected result, disclosing the destructive interference of the massive mode\cite{CW}
resulting from the simultaneous soldering of dynamics and symmetry.

The main purpose of this investigation is to recover the dynamics
of the chiral interference.  To this end
we introduce, in section III,
the new idea of {\it dynamical decomposition} that separates
the chiral sector from the symmetry sector in the Siegel CBS. 
Under this field redefinition, the Siegel action can
be reexpressed as a FJ action carrying the chirality, plus a noton,
carrying the representation of the Siegel symmetry.
This proposal shows that the difference between these
CBS's is given by the presence of a noton.
We stress that the noton is a nonmover field at classical level that
acquires dynamics upon quantization.  To find out that the noton is
already contained in the Siegel's theory is a new and outstanding
result. We end up the section examining the contribution of the noton
to the coefficient of the Schwinger term
in the energy-momentum tensor current algebra, where it is shown that
its quantum dynamics is fully responsible for the Siegel anomaly.
With the dynamical and the symmetry sectors isolated,
we might introduce independent soldering
to avoid the destructive interference of the chiral modes. This is
done in section IV.
In the last section we discuss the importance and consequences of our findings.


\noindent{\bf 2}  In this section we study the interference of gauged rightons and leftons
in the framework of \cite{MS}\footnote{The results of 2D-soldering may, equivalently, be obtained
including a right-left fermionic interaction term, before bosonization, as shown first in \cite{DGR}.}.
Gates and Siegel\cite{GS} have examined the interactions of leftons and
rightons with external gauge fields including the supersymmetric and the
nonabelian cases.  Let us write the
Siegel action with an electromagnetic coupling\cite{GS} as,

\begin{equation}
\label{sie1}
S_{\mp}^{(0)} = <\partial_{\pm}\phi_{\pm}(\partial_{\mp}\phi_{\pm} + 2A_{\mp}) +
\lambda_{\pm\pm} 
(\partial_{\mp}\phi_{\pm}+A_{\mp})^{2}>.
\end{equation}
Here $\phi_{\pm}$ describes a lefton (righton) and $<...>$ means space-time integration. 
This action has been used in \cite{GS}
to bosonize a generalized Thirring model that was shown to be invariant
under the extended conformal transformation,
\begin{eqnarray}
\label{II1}
\delta \, \phi_{\pm} & = & \xi^\mp\,(\,\partial_\mp \phi_{\pm} \; + \; A_\mp) \nonumber \\
\delta \lambda_{\pm\pm} & = & -\,\partial_\pm\,\phi_{\pm} \;+\; 
\xi^\mp\,\stackrel{\leftrightarrow}{\partial}_\mp\,\lambda_{\pm\pm}\nonumber\\
\delta A_\mp &=& 0 \;\;.
\end{eqnarray}

\noindent In the soldering formalism one must examine
the response of the system to the transformation\cite{MS,ABW} 
\begin{equation}
\label{II2}
\phi_\pm \rightarrow \phi_\pm + \alpha
\end{equation}
and compute the corresponding Noether currents,
\begin{eqnarray}
\label{II3}
J_{\phi_\pm}^{\mp} = 2 \, [ \,\partial_{\pm}\phi_\pm \;+\; \lambda_{\pm\pm}\,
(\partial_{\mp}\phi_\pm \;+\; A_{\mp}) \,];\;\;\; J_{\phi_\pm}^{\pm} = 2 \, A_{\mp}.
\end{eqnarray}

\noindent Following \cite{MS} and \cite{ABW}, we construct an
iterated action introducing the soldering field $B_\mu$,
\begin{equation}
\label{a1}
S_{\mp}^{(0)} \rightarrow  S_{\mp}^{(1)} = S_{\mp}^{(0)} -
<B_{\mu}\,J_{\phi_\pm}^{\mu}>+<\lambda_{\pm\pm} \, B^{2}_{\mp}>.
\end{equation}
This iterated action behave, under the axial gauge transformation (\ref{II2}) and, $\delta B_{\pm} = \partial_{\pm}\alpha$ as,

\begin{equation}
\label{var}
\delta S_{\mp}^{(1)} \;=\; - \, 2 \, <B_{\mp} \, \delta B_{\pm}>\;\;.
\end{equation}
We can see that $S^{(1)}_\mp$ are not invariant, but their variations
are independent of the chiral fields $\phi_\pm$ and the result
(\ref{var}) reflects the anomalous behavior of the chiral model
under gauge transformations.  However,  being dependent only on the fields
taking values on the gauge algebra, they might cancel out mutually.
Indeed, the sum of $S^{(1)}_+$ and $S^{(1)}_-$
together with a contact term of the form $2\,<B_{-}\,B_{+}>$
results in an invariant action, 
\begin{eqnarray*}
\lefteqn{S_{T}  =  <\partial_{+}\phi \, (\,\partial_{-}\phi +2A_{-}\,) 
+\lambda_{++} \, (\,\partial_{-}\phi+A_{-}\,)^{2}} \\ 
& & +  \partial_{-}\rho \, (\,\partial_{+}\rho +2A_{+}\,)+
 \lambda_{--} \, (\,\partial_{+}\rho+A_{+} \, )^{2}
- \; B_{\mu}\,J_{\phi}^{\mu}\\
& &   - B_{\mu}\,J_{\rho}^{\mu}+ \lambda_{--}\,B_{+}^{2} +  \lambda_{++}\,B_{-}^{2} + 
2\,B_{-}\,B_{+}>
\end{eqnarray*}
where we have used $\phi_+ = \phi$ and $\phi_- =\rho$ for clarity.  Eliminating $B_\mu$ from their field equations,
leads to an effective action that incorporates the effects of the
right-left interference
\begin{eqnarray}
\label{II11}
\lefteqn{S_{eff}  = <\Delta\left\{\left(1 + \lambda_{++}\lambda_{--}\right) 
\partial_{-}\Psi \partial_{+}\Psi\right.} \nonumber\\ 
&& \left. + \lambda_{++}\left(\partial_{-}\Psi\right)^2 + 
\lambda_{--}\left(\partial_{+}\Psi\right)^2\right\} > - 2<A_{-}A_{+}>
\end{eqnarray}
\noindent where $\Delta =\left(1- \lambda_{++}\lambda_{--}\right)^{-1}$ and
$\Psi = \rho - \phi$.  The soldering process is now completed.  We have succeeded in including the effects of
interference between rightons and leftons.  Consequently, these components have lost their
individuality in favor of a new, gauge invariant,
collective field  that 
does not depend on $\phi$ or $\rho$ separately. 

The physical meaning of (\ref{II11}) can be appreciated by
solving for the multipliers and using the symmetry induced by the soldering\cite{CW}, showing that it represents the action for the
noton.  In fact (\ref{II11}) is basically the action
proposed by Hull\cite{CH} as a candidate for canceling the Siegel
anomaly. This field carries a representation of the full diffeomorphism group\cite{CH}
while its chiral (Siegel) components carry the
representation of the chiral diffeomorphism.
Observe the
complete disappearance of the dynamical sector due to the destructive interference between the leftons and the rightons.  This happens
because we have introduced only one soldering field to deal with both the dynamics and the symmetry.
To recover dynamics we need to separate these sectors and solder them
independently.  This we do in the next section.


\noindent {\bf 3}  The main result of this paper is presented in this section. 
We propose that different CBS's are related by the presence of a noton. 
We show that a Siegel mode
may be decomposed in a FJ mode, responsible for the dynamics, and a 
nonmover, carrying the representation
of the symmetry group.  This is done introducing a {\it dynamical
redefinition} in the phase space of the model.
We stress that these fields are independent as they originate from completely different actions, and the presence of a Siegel noton was not
pointed in the literature so far.
This new result is complementary to the established knowledge, where the FJ action is interpreted as a gauge fixed Siegel action\cite{WS}. 
Under this new point of view we look at the gauge fixing process as the condition that sets the noton field to vanish.
These points will now be clarified.
   
Let us begin with the Siegel action for a left-mover scalar,
\begin{equation}
\label{1order}
S = <\pi\,\dot{\phi}- \frac{{\phi'}^2}{2} -\frac{1}{2} \frac {\left(\pi-\lambda_{++}{\phi'}\right)^2}{1-\lambda_{++}}
-\frac{\lambda_{++}}{2}{\phi'}^2>.
\end{equation}
One can fix the value of the multiplier as $\lambda_{++} \rightarrow 1$ to reduce
it to its FJ form. The phase space of the model is correspondingly reduced to $\pi \rightarrow \phi'$.
The third term in (\ref{1order}) reduces to $(1-\lambda_{++}){\phi'}^2\rightarrow 0$
as $\lambda_{++}$ approaches its unit value.
This reduces the symmetry of the model, leaving behind
its dynamics described by a FJ action.
The above behavior suggests the following field redefinition,
\begin{eqnarray}
\label{theresult}
\phi = \varphi + \sigma;\;\;\;\pi = \varphi' - \sigma',
\end{eqnarray}

\noindent which is our cherished result.  The lefton $\phi$ is
related to the FJ chiral mode $\varphi$ by the presence of a noton
$\sigma$. Such a decomposition immediately diagonalizes (\ref{1order}) as,
\begin{equation}
\label{III2}
S  =  <{\varphi'}\,\dot{\varphi}\;-\;{\varphi'}^2>\;+\;
 <-\;\sigma'\dot{\sigma}\;-\;\eta_+\,\sigma'^2>\;\;,
\end{equation}
and $\eta_{\pm} = \frac{1\;\;+\;\;\lambda_{\pm\pm}}{1\;\;-\;\;\lambda_{\pm\pm}}$.  In this form, the chiral information is displayed by the FJ field ${\varphi}$ while the noton $\sigma$ carries the symmetry of the original model.
The reduction of the phase space is attained by letting the noton
$\sigma$ approach zero as the multiplier $\eta_+$ diverges.  This eliminates the symmetry carrying sector leaving behind only the FJ mode.  

To disclose the meaning of the symmetry, we need to study the noton invariances, imposed by the constraint $\sigma^2\approx 0$.  Following sympletic formalism\cite{annals}, we obtain  from (\ref{III2}), the following sympletic matrix   
\begin{equation}
\label{III4}
f = 2 \left( \begin{array}{cc}
               1 & {\sigma_y}' \\
              {\sigma_x}' & 0
             \end{array} \right)\;\delta'(\,x\,-\,y\,)
\end{equation}
whose single zero-mode displays the searched symmetry. 
\begin{eqnarray}
\label{III5}
\delta\,\sigma  =  \epsilon\,\sigma';\;\;\;
\delta\,\eta  =  \dot{\epsilon} + \eta\,\epsilon' - \epsilon\,\eta'.
\end{eqnarray}
As claimed, $\sigma$ carries a representation of the Siegel algebra.  Solving the equations of motion
and making use of (\ref{III5}), we find that $\sigma$ is indeed a nonmover.  
In this way we have realized, in a deeper level, the decomposition
of the Siegel's chiral boson in terms of dynamics and symmetry
and identified the associated fields.

When the quantization process is accomplished the noton acquires dynamics through the gravitational anomaly\cite{Quant}.  To see this we examine its quantum contents and show that it 
contributes fully to the Siegel anomaly.
This is done by computing the Schwinger terms of the energy-momentum
tensor current algebra in the noton action.
We begin with a separation of the field operator in
creation and annihilation parts\cite{Fuji}, $A = A_+\, + \,A_-$
and define the vacuum state by $A_-\,|0> = 0$.
Define the positive and negative frequency projector as,
\begin{equation}
\label{III8}
A_{\pm}\,(x) = \int\,dz\,\delta_{\pm}\,(x-z)\,A(z)
\end{equation}

\noindent where the chiral delta functions are defined as
\begin{equation}
\label{III10}
\delta_{\pm} \,(x)= \mp\, {i\over{2\,\pi}}{1\over{x\,\mp\,i\,\epsilon}}
\end{equation}

\noindent and satisfy the following property,
\begin{equation}
\label{III11}
\left(\,\delta'_{+}\,(x)\,\right)^2 - \left(\, \delta'_{-} \,(x)\right)^{2} = {i\over{12\pi}} \delta'''\,(x) .
\end{equation}

\noindent To compute the Schwinger term, we examine the energy-momentum tensor $T\,(x) = [\,\sigma'(x)\,]^2$,
whose classical algebra is,
\begin{equation}
\label{III15}
\{\,T(x),T(y)\,\} = ( \,T(x)\,+\,T(y)\,)\,\delta'(x\,-\,y)
\end{equation}


\noindent Then, upon quantization, the presence of a Schwinger
term\cite{jackiw} is completely disclosed by normal ordering
the energy-momentum operator. Call $\sigma'(x)\,=\,\xi(x)$,
with $\xi_+$ and $\xi_-$
being the creation and annihilation operators respectively.
Using that,
\begin{equation}
\left[\,\xi(x)\,,\,\xi(y)\,\right] \;=\; \frac{i\,\hbar}{2}\,\delta'(\,x\,-\,y\,),
\end{equation}

\noindent it is easy to check that the following results are obeyed
\begin{eqnarray}
\left[\,\xi_{\pm}(x)\,,\,\xi(y)\,\right] \;&=&\; \frac{i\,\hbar}{2}\,\delta'_{\pm}(\,x\,-\,y\,),\nonumber\\
\left[\,T(x)\,,\,\xi_{\pm}(y)\,\right] \;&=&\; i\,\hbar\,\xi(x)\,\delta'_{\mp}(\,x\,-\,y\,)\;\;.
\end{eqnarray}

\noindent The current-current commutator,
\begin{eqnarray}
& &{\left[\,T(x)\,,T(y)\right] = i\,\hbar\,\left(T(x)+T(y)\right)\,\delta'\,(\,x\,-\,y\,)}\nonumber  \\
& & \mbox{}+ \frac{i\,\hbar^2}{24\,\pi}\,\delta'''\,(\,x\,-\,y\,)
\end{eqnarray}
is identical to the well known current algebra for the energy-momentum tensor operator for the Siegel model\cite{Son} with the correct value
for the central charge.  This shows that the noton 
present in the Siegel formulation completely takes care of the symmetries,
both classically and quantically.  This is a new and outstanding result.  It explains why the Hull's mechanism for canceling
the Siegel anomaly works by including a properly normalized external noton.


\noindent {\bf 4}  We are now in position to apply the dynamical decomposition to
the actions (\ref{sie1}), and study the effect of their interference.
According to the discussion of the last section
these systems decompose in chiral, right and left,
scalars $\varphi$ and $\varrho$ coupled to gauge fields,
and two notons, $\sigma$ and $\omega$, carrying the
representation of the right and the left Siegel symmetry,

\begin{eqnarray}
S_{-}^{(0)} & = & <\left\{ {\varphi'}\,\dot{\varphi}-{\varphi'}^2+2\,\sqrt{2}\,A_{-}\,{\varphi'}-
A_{-}^2+ {a \over 2}\,A_{\mu}^2 \right\}> + \nonumber\\
& + & <\left\{ -{\sigma'}\,\dot{\sigma}-\eta_{+}{\sigma'}^2\right\}>\nonumber \\
S_{+}^{(0)} & = & <\left\{-{\varrho'} \dot{\varrho}-{\varrho'}^2-2\,\sqrt{2}\,A_{+}\,{\varrho'}-A_{+}^2+ 
{b \over 2}\,A_{\mu}^2\right\}> + \nonumber\\
& + &<\left\{\;{\omega'}\,\dot{\omega}-\eta_{-}{\omega'}^2\right\}>.
\end{eqnarray}

\noindent The coefficients $a$ and $b$ are the Jackiw-Rajaraman regularization
parameters.  Each sector, dynamics or symmetry, corresponds to a self-dual or antiself-dual aspect
of the chirality.  This sets the stage for the independent soldering.  Consider the behavior
of the above theories under the following axial gauge transformation,
\begin{eqnarray}
\left(\varphi, \varrho\right)\;\; & \longrightarrow &\;\;
\left(\varphi\;\;+\;\; \alpha, \varrho \;\;+\;\; \alpha\right)
\nonumber \\
\left(\sigma, \omega\right)\;\; & \longrightarrow & \;\;
\left(\sigma \;\;+\;\; \beta, \omega \;\;+\;\; \beta\right)
\end{eqnarray}
which introduces two independent soldering.  The corresponding Noether's currents,
${\cal J}_{\alpha}^{\mu(\pm)}$, with $\mu=0,1$ being the Lorentz index and $\pm$ indicating the chirality,  are,
\begin{eqnarray}
{\cal J}_{\alpha}^{0(+)} & = & {\cal J}_{\alpha}^{0(-)} = 0 \nonumber \\
{\cal J}_{\alpha}^{1(+)} & = & -\;\;2\,(\,\dot\varrho \;\;+\;\;\varrho' \;\;+\;\;\sqrt{2}\,A_+\,) \nonumber \\
{\cal J}_{\alpha}^{1(-)} & = & 2\,(\,\dot{\varphi} \;\;-\;\;{\varphi'} \;\;+\;\;\sqrt{2}\,A_-\,)
\end{eqnarray}
and ${\cal J}_{\beta}^{\mu(\pm)}$
\begin{eqnarray}
{\cal J}_{\beta}^{0(+)} & = & {\cal J}_{\beta}^{0(-)} = 0 \nonumber \\
{\cal J}_{\beta}^{1(+)} & = & 2\,(\,\dot\omega \;\;-\;\;\eta_-\,\omega'\,) \nonumber \\
{\cal J}_{\beta}^{1(-)} & = & -\;\;2\,(\,\dot\sigma \;\;+\;\;\eta_+\,\sigma'\,) \;\;.
\end{eqnarray}
Following Refs.\cite{MS,ABW}, we get, after double iteration and elimination of the auxiliary vector soldering fields, the following effective action
\begin{equation}
S_{eff} = S_{+}^{(0)} \;\;+\;\;S_{-}^{(0)} \;\;+\;\; \frac{{\cal J}^2_{\alpha}}{8} \;\;+\;\; \frac{{\cal J}^2_{\beta}}{8\,\eta} 
\end{equation}
where
\begin{eqnarray}
\eta & = & \frac{\eta_- \;\;+\;\; \eta_+}{2} \nonumber \\
{\cal J}_{\alpha} & = &  {\cal J}_{\alpha}^{1(+)}+{\cal J}_{\alpha}^{1(-)} \nonumber\\
{\cal J}_{\beta} & = &  {\cal J}_{\beta}^{1(+)}+{\cal J}_{\beta}^{1(-)}\;\;.
\end{eqnarray}
After a tedious algebraic manipulation, we see that the final form of the action is independent of the individual fields, depending only on their gauge invariant combinations $\Phi$ and $\Psi$,
\begin{eqnarray}
S_{eff} & = &<\left\{ \frac{1}{2}\,\partial_{\mu}\,\Phi\,\partial^{\mu}\,\Phi+2\,\epsilon^{\mu \nu}\,A_{\mu}\,\partial_{\nu}\,\Phi 
+ \eta_{0}\,A_{\mu}^2\right\}>+ \nonumber \\
& + & <\left\{\frac{1}{2\,\eta}\,\dot{\Psi}^{2}-\frac{\eta_-\,\eta_+}{2\,\eta}\,{\Psi'}^2+\,\eta_{1}\dot{\Psi}\,\Psi'\right\}>,
\end{eqnarray}
where $\eta_{0}= \frac{1}{2}\left(a+b-2\right)$, 
$\eta_{1} = \frac{\eta_{+} -\eta_-}{2\,\eta}$ and the new collective fields are $\Phi = \varphi - \varrho$ and $\Psi =\sigma - \omega$.

In the case where $a=b=1$, the first part of this action leads to the Schwinger model.  The combination of the massless modes led to a massive vectorial mode as a consequence of the chiral interference. The noton action was shown to propagate neither to the left nor to the right\cite{CH}. This action has the
same form as the one defined by Hull to cancel the  Siegel anomaly.  Noticeably, this noton has not coupled to the gauge field of the theory.

The separation of the symmetry sectors (notons) followed by soldering has led to another new and interesting result.  The dynamical chiral sectors interfered to produce a vector massive mode\cite{ABW}. The Siegel notons, on the other hand, led to a Hull noton.  However, while each component carry a representation of the Siegel algebra, the soldered (Hull) noton carries a representation of the full diffeomorphism group.  It is worth to stress that Siegel invariance is not a diffeomorphism sub-group so that the soldering gives more than the mere direct sum of the chiral algebras \cite{DGR}.


\noindent {\bf 5}  The chiral bosonization process is not free of ambiguities.  This is different from the vectorial case, where the presence of the gauge symmetry leads to an exact and unambiguous result.  In particular, it is well known that the bosonization dictionary,
where  $\bar\psi i\partial\!\!\!/\psi\rightarrow\partial_+\phi\partial_-\phi$  
and  $\bar\psi\gamma_\mu \psi\rightarrow\frac{1}{\sqrt{\pi}}
\epsilon_{\mu\nu}\partial^\nu \phi$  cannot be applied in the chiral case.  This ambiguity has been characterized by Jackiw and Rajaraman\cite{JR} with an arbitrary mass parameter regulator.  To pass from the vectorial to the chiral case one has to destroy the right-left interference.  As seen in \cite{ABW} the interference contributes to the mass term.  Therefore its absence
leaves the mass coefficient arbitrary.  This stands at the origin of the Jackiw-Rajaraman effect.

The arbitrariness in the process of chiral bosonization has led to different bosonization schemes\cite{WS,FJ,many}.  In this work we compared the two most successful proposals\cite{WS,FJ}, and showed that they are related by the presence of a noton.  We proved that the dynamical decomposition diagonalizes Siegel action into a dynamical and a symmetry carrying parts.  This gives a deeper insight in the composition of the Siegel mode.  We have also analyzed the quantum contents of the noton sector and showed that although it is a nonmover field classically, it acquires dynamics at the quantum level thanks to the gravitational
anomaly.

Throughout this paper we made use of the soldering formalism.  In this context the dynamical decomposition was important to clarify the reason why gauged rightons and leftons fail to produce a constructive interference pattern. Separating the symmetry carrying notons before soldering, a double constructive interference was made possible.  This process led to a massive vector mode in the dynamical sector.
The symmetry sector is described by a Hull noton, carrying the representation of the full diffeomorphism group.
It was the soldering
that allowed for the construction of the full group
in terms of its chiral parts by incorporating the interference term.

To conclude, we would like to speculate on possible uses of the above result.  It would be interesting to examine if the alternative fermionic noton\cite{DGR} also used to cancel the anomaly is able to produce a SUSY pair with the Siegel noton.
It seems equally interesting to examine if the dynamical decomposition that allowed for the separation of dynamics and symmetry works in self and
anti-self dual systems in dimensions different from two.  Finally, it seems suggestive that the soldering might lead to new and interesting combinations of algebras other than the diffeomorphism.

\noindent {\bf Acknowledgment:} This work is supported in part by
CNPq, CAPES, FINEP and FUJB (Brazilian Research Agencies).

\end{document}